\documentclass[osajnl,twocolumn,superscriptaddress,11pt]{revtex4-1} %% use 11pt for Applied Optics
\usepackage{amsmath,amssymb,graphicx}
\usepackage{setspace}

\begin{document}
\singlespacing

\title{Intensity distribution of the parhelic circle and embedded parhelia at low solar elevations: theory and experiments}

\author{Sarah Borchardt}
\affiliation{Universit\"at Leipzig, Linn\'estr.\ 5, 04103 Leipzig, Germany}
\author{Markus Selmke}\email{Corresponding author: selmke@rz.uni-leipzig.de}
\affiliation{Universit\"at Leipzig, Institute of Experimental Physics I, Molecular Nanophotonics Group, Linn\'estr.\ 5, 04103 Leipzig, Germany}

\begin{abstract}
We describe the intensity distribution of the parhelic circle for plate-oriented hexagonal ice crystals at very low solar elevations using geometrical optics. An experimental as well as theoretical study of in-plane ray-paths provides details on the mechanism for several halos, including the parhelia, the $120^{\circ}$ parhelia, the blue spot and the Liljequist parhelia. Azimuthal coordinates for associated characteristic features in the intensity distribution are compared to data obtained using a rotating hexagonal glass prism.
\end{abstract}

%\ocis{(010.1290) Atmospheric optics; (010.2940) Ice crystal phenomena; (010.1690) Color; (010.4030) Mirages and refraction.}% REPLACE WITH CORRECT OCIS CODES FOR YOUR ARTICLE
                          % NOTE: \ocis{} IS ALIASED TO \pacs{} BUT MUST
                          % FORMAT THE TERMS CORRECTLY FOR EACH JOURNAL

\maketitle %% required

\section{Introduction}
Atmospheric optics phenomena associated with refraction represent instructive examples for the concepts of geometrical optics. The underlying principles of the formation of a rainbow for instance provide a classic example in standard courses of optics. Table-top demonstrations include the illumination of a round-bottomed flask containing water, a single suspended droplet of water \cite{Walker1975} or a transparent acrylic cylinder \cite{Casini2012}. The much lesser-known phenomena of parhelia (PH), also called son dogs, are attributed to hexagonal plate-like ice-crystals instead of water drops. They can be observed all-year and even more frequently than rainbows by the informed observer: On average, they can be seen twice a week in parts of the United States and Europe \cite{TapeBook,atmosHP} and only require high clouds or cold air harbouring the tiny crystals. At low solar elevation, these crystal refract light by an angle no less than $\theta_{\rm PH}^{\rm min} = 22^{\circ}$, corresponding to the azimuth at which the parhelia appear. But then again, this particular halo is only the most common type of a whole panopticon of intricate arcs, bows and spots that occasionally decorate the sky. Apart from chemical methods to generate artificial halos that have been at least as early as 1831 \cite{BrewsterTreatise,Cornu1889,Tammer1998}, the classic rotating prism experiment has long been used since A.\ Bravais \cite{Bravais1847,BravaisMemoir,Wegner1917} to study halos, commonly using equilateral prisms. Several more recent publications in the English literature \cite{Berry1997,Tammer1998,Greenler2003} reported on such single crystal laboratory demonstrations, showcasing their potential in understanding ice halo phenomena. In contrast to equilateral prisms, hexagonal glass prisms provide a fair analog of the hexagonal ice crystals which produce halos. As homogenizing light pipes \cite{Zhu2013}, high precision glass prisms with hexagonal cross-section have recently become commercially available (e.g.\ Edmund Optics GmbH). The artificial halos which can be investigated include several that cannot be produced with triangular prisms \cite{Tammer1998,Greenler2003}.

Although the index of refraction of the common BK7 glass prism material, $n=1.530$ for blue light and $n=1.515$ for red light, is different from that of ice of about $n= 1.31$, the dispersion is similar. Longer wavelengths, i.e.\ red colors, experience a slightly lower index of refraction than shorter wavelengths (blue colors). Consequentially, the coloring of the halos, artificial in an experiment or natural, is similar. Still, the angles at which they appear and the extent for each halo will differ. %\cite{RefractiveIndexWP}

%
%\begin{figure}[b]
%\begin{center}\includegraphics [width=1.0\columnwidth]{Sketch_SundogFormation.pdf} %figure1.png
%  \end{center}\caption{Typical display of the two parhelia (sundogs) for horizontally oriented plate-like crystals. The parhelic circle is often too faint to be observed, as in this case. The $22^{\circ}$ halo is caused by the same light paths but for randomly oriented plate and column crystals.\label{FigPH}}
%\end{figure}
%

%\begin{figure}[b]
%\centerline{\includegraphics [width=1.0\columnwidth]{Fig_nthetaPlots.pdf}} %figure1.png
%\caption{Dispersion curve for the index of refraction of ice and BK7 glass.\label{Fig_ntheta}}
%\end{figure}

Combined with a rotary stage, a single hexagonal rod can simulate the random orientations about the vertical axis in a large ensemble of plate-oriented crystals that cause halos such as parhelia \cite{TapeBook}. This allows a detailed and instructive investigation of a series of phenomena using only ray-optics supplemented by Fresnel's transmission and reflection coefficients. In particular, the parhelic circle (PHC) becomes apparent, which is a circle at the solar almucantar that spans the entire azimuthal range of the sky \cite{Bravais1847,TapeBook,Lynch1978,atmosHP}. Three different types of crystal populations can produce this halo, including plate oriented, horizontal columns and Parry oriented crystals. The former one is typically the most common contributor. We therefore focus on plate oriented crystals, both experimentally and theoretically. The PH may be considered as being just the most prominent contributor to the PHC intensity pattern. The common $120^{\circ}$ parhelia, the recently confirmed \cite{BlueSplot2001} blue spot and the Liljequist parhelia which appear at low sun elevations are further features of the comparatively complex PHC. The ray paths that contribute to the PHC range from simple external reflection to more intricate paths involving two refractions and several internal reflections \cite{Bravais1847,TapeBook,atmosHP}. As we will show, prism experiments and ray-by-ray studies permit a rich understanding of the contributing mechanisms, at least for near-zero solar elevations.

\section{Ray optics and intensities}
\begin{figure}[t]
\centerline{\includegraphics [width=1.0\columnwidth]{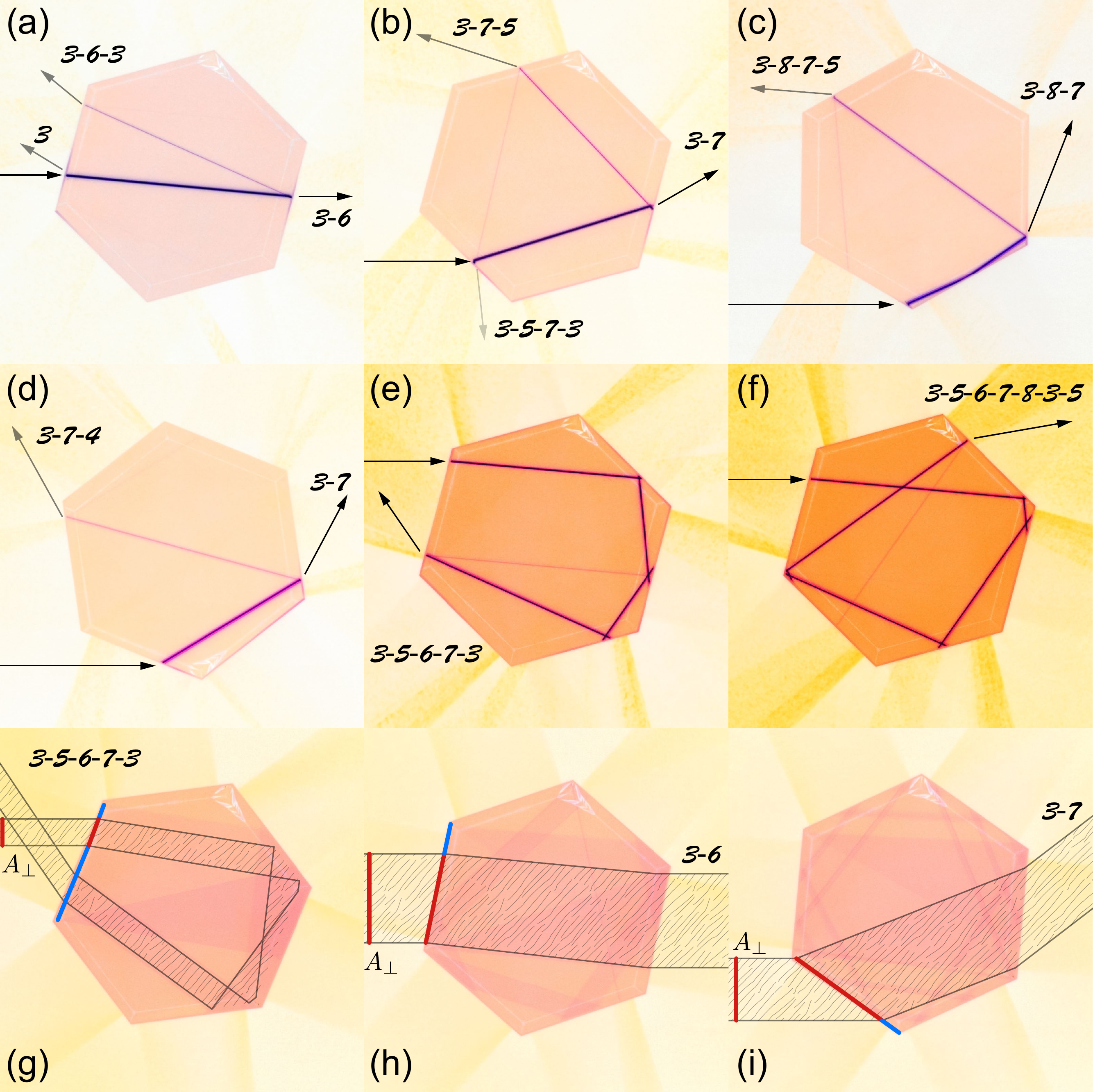}} %figure1.png
\caption{Color-inverted (false colour) photographs of various light rays through a BK7-hexagon using a focusable laser-diode. \textbf{(a)-(f)} Consecutive segments of equal intensity signify total internal reflection, whereas a change indicates a partial reflection according to the Fresnel coefficients. Rays of type $\mathfrak{3}\textnormal{-}\mathfrak{7}$ are responsible for the $22^{\circ}$-PH, $\mathfrak{3}\textnormal{-}\mathfrak{5}\textnormal{-}\mathfrak{7}\textnormal{-}\mathfrak{3}$ rays for the $90^{\circ}$-PH, and $\mathfrak{3}\textnormal{-}\mathfrak{8}\textnormal{-}\mathfrak{7}\textnormal{-}\mathfrak{5}$ as well as $\mathfrak{3}\textnormal{-}\mathfrak{5}\textnormal{-}\mathfrak{6}\textnormal{-}\mathfrak{7}\textnormal{-}\mathfrak{3}$ for the Liljequist PH. \textbf{(g)-(i)} Defocused parallel illumination shows several bundles of rays each following the same path (bottom row). \label{FigPaths}}
\end{figure}

Using the laser beam of a focusable laser diode (cheap ones are available on ebay), the various types of ray paths that occur for a hexagonal prism can be investigated individually. Specifically, using a blue laser diode has the distinct advantage of a visible ray path throughout the prism via the excited autofluorescence of the prism material. Some of the most important rays for the PHC, excluding rays entering the top face, are shown in Fig.\ \ref{FigPaths}. A discussion of these requires some classification. We here adopt the system of the book by Walter Tape \cite{TapeBook}: A ray path is notated according to the order in which it encounters the prism faces. The top and bottom basal faces are numbered $\mathfrak{1}$ and $\mathfrak{2}$, respectively, while the remaining side faces are numbered $\mathfrak{3}$ to $\mathfrak{8}$ in counter-clockwise fashion. Still, each ray will possess a mirror ray, but both will be named after the member which gives rise to an intensity towards the right side of the sun. A longer string corresponds to a more complex path.

Theoretical investigations of halo characteristics using geometrical optics \cite{Liou1981,Takano1985,Greenler1986,Liou1989,TapeBook,Macke1996,Borovoi2007} can be done using any of the freely available halo simulation tools. HaloSim \cite{HaloSim} for instance allows the refractive index to be set, and choosing a user defined fisheye perspective centering on the zenith allows simulation of the entire sky, including the full PHC. Details on polarisation and additional diffraction \cite{Flatau2014} may for small crystals also be included in such models. However, while such software packages give a good impression of the whole sky and a wealth of information on the observable halos, the Monte Carlo approach employed does not lend itself to analytical progress concerning the angular positions of halo features.

\begin{figure}[b]
\begin{center}\includegraphics [width=1.0\columnwidth]{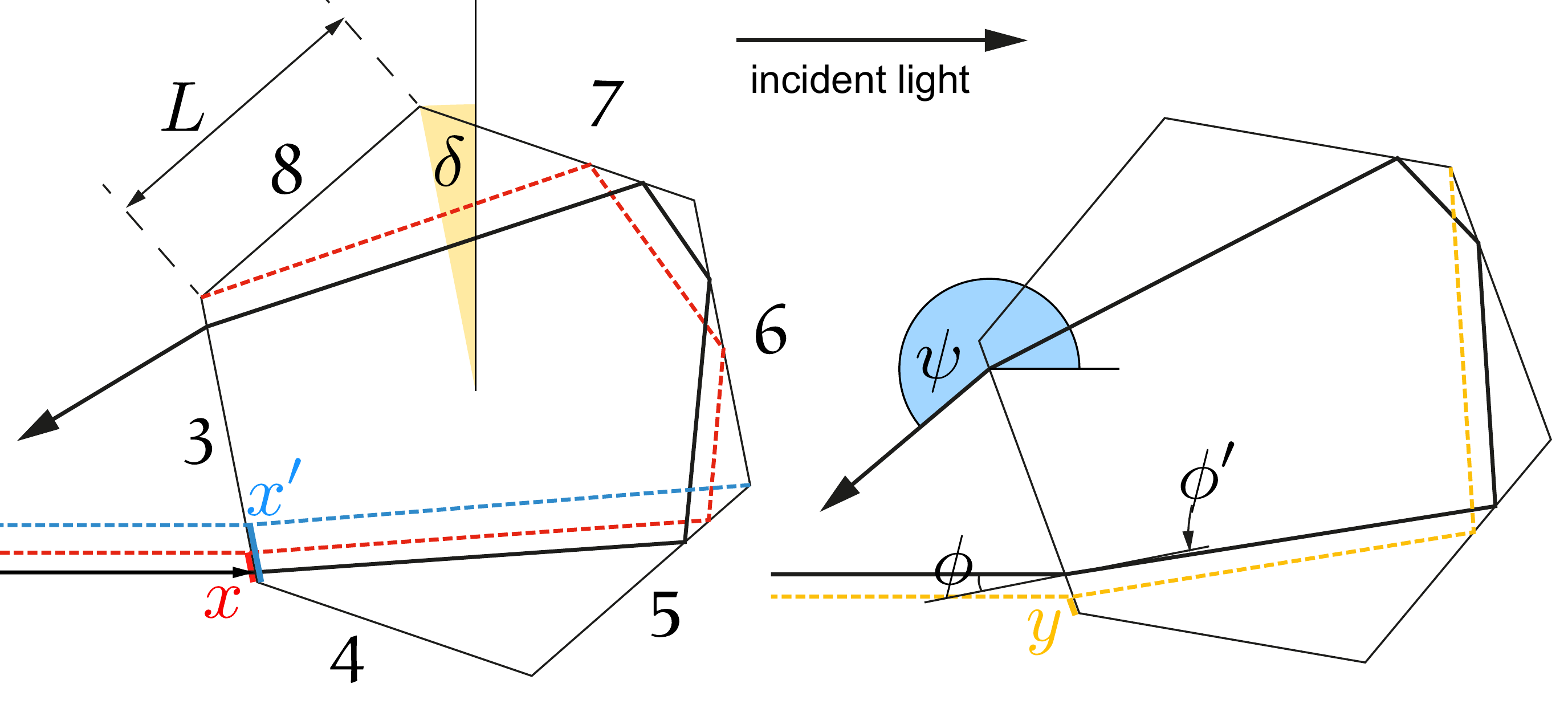} %figure1.png
  \end{center}\caption{Geometry for the geometrical optics calculation: $\delta$ is the prism angle, $\psi$ the total deflection angle, $\phi$ the incident angle and $\phi'$ the refracted internal angle, both related by $\sin\left(\phi\right)=n\sin\left(\phi'\right)$. The sketched ray (black) is one of two possible $\mathfrak{3}\textnormal{-}\mathfrak{5}\textnormal{-}\mathfrak{6}\textnormal{-}\mathfrak{7}\textnormal{-}\mathfrak{3}$ paths for $\delta\in\left[0,30^{\circ}\right]$. Its limiting ray paths (dashed) determine the lengths $x=\left[-1+5\sqrt{3}\tan\left(\phi'\right)\right]L/2$, $x'=\sqrt{3}\tan\left(\phi'\right)L$ and $y=\left[-1+3\sqrt{3}\tan\left(\phi'\right)\right]L/2$. $y$ and $x$ were obtained by successive application of the law of sines. For this entry face the incidence angle is $\phi=\delta$. The effective area is then $A_{\perp}^{(m)}=\left[{\rm max}\left(0,y\right) - {\rm max}\left(0,{\rm min}\left(x,x'\right)\right)\right]\cos\left(\phi\right)$.\label{FigArea}}
\end{figure}

A simpler and more insightful approach is suitable for the PHC due to plate oriented crystals at low solar elevations. It will therefore be particularly appropriate for the experimental situation of a rotating prism. A detailed analysis of the ray paths and their geometrical constraints was outlined in Refs.\ \cite{White1976,Lynch1978Anthelion,Denny1997}. Extending this approach to all major ray-paths of the PHC, analytical expressions can be found for those azimuthal angles at which characteristic intensity features within the PHC appear, and the range of certain contributions can be given explicitly. The by-hand method further permits more insight to be gained about the factors that determine the intensity pattern. Firstly, partial transmittance and reflection occur at the interfaces according to the squared Fresnel amplitude ratios $r$ (polarisation averaged). Depending on the incidence angle for each face encountered, a ray may also experience total internal reflection whenever the critical angle $\alpha_{\rm TIR}=\arcsin\left(1/n\right)$ is exceeded. The laser-beam's path through the glass prism shown in Fig.\ \ref{FigPaths}(a)-(f) makes both phenomena directly visible: Two consecutive segments of the laser ray, i.e.\ before and after hitting a side face, will either show different or equal intensities depending on the particular incidence angle. Secondly, geometrical constrains limit the effective area $A_{\perp}$ intercepted by the hexagonal crystal's side faces, typically being less than the full face of length $L$ (times a unit height). Each path allows only a certain pencil of rays to emerge from the exit face \cite{TapeBook}, as illustrated in Fig.\ \ref{FigPaths}(g)-(i). This effective area can be found by geometrical means as illustrated exemplarily for a Liljequist PH ray-path in Fig.\ \ref{FigArea}.

The intensity $I\left(\theta\right)$ due to all $m$ considered ray-paths (each involving at most $k$ internal reflections) can be found by binning the total deflection angles $\psi_{(m)}\left(\delta\right)$ for all paths at varying prism orientations $\delta\in\left[0,30^{\circ}\right]$ \cite{Denny1997}. Each orientation and path then adds to the intensity an amount
%The intensity for all $m$ considered ray paths with at most $k$ reflections and refractions then reads
%
\begin{equation}
\mathrm{d} I\left(\theta\right)\propto \frac{A_{\perp}^{(m)}\left(\delta\right)}{2L}\prod_{i=1}^{k}r_{(m,i)}^2\left(\delta\right),\label{eq:IntGOA}
\end{equation}
multiplied with the appropriate transmittance through the entrance and exit faces, i.e.\ two factors of the form $\left[1-r^2\right]$. Deflection angles lying within the left and right hemisphere are taken to contribute to the same intensity within $\theta\left(\psi_{(m)}\right)\in \left[0-180^{\circ}\right]$. Therefore, the pattern already repeats for prism-orientations $\delta > 30^{\circ}$. Accordingly, only 3 faces need to be considered as possible entry faces \cite{Denny1997}. For instance, the intensity distribution of the type $\mathfrak{3}\textnormal{-}\mathfrak{7}$ rays was computed via rays entering the three left-most faces in Fig.\ \ref{FigArea} (faces $\mathfrak{3},\mathfrak{4},\mathfrak{8}$). The supplementary material includes a map containing all paths and their deflection angles as a function of the prism orientation, as well as the limiting ray paths necessary to find the effective areas $A_{\perp}^{(m)}$. An open-source real-time implementation of the code is available upon request. 

\begin{figure}[b]
\centerline{\includegraphics [width=1.0\columnwidth]{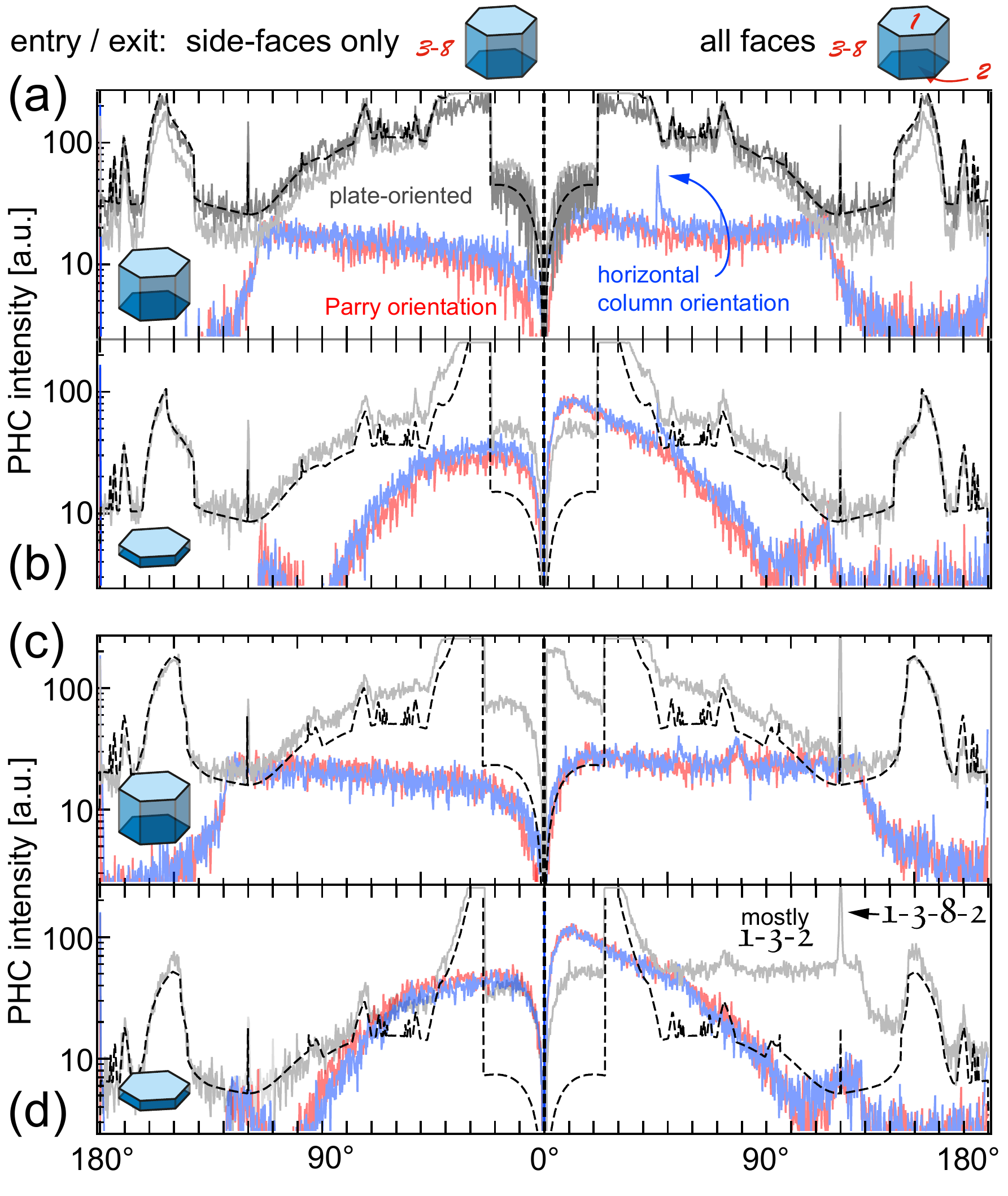}}
\caption{PHC intensity distribution for light source elevations \textbf{(a),(b)} $e=1^{\circ}$ and \textbf{(c),(d)} $e=20^{\circ}$. The simplified model is shown as dashed lines. The noisy lines show a full Monte Carlo simulation ($5\times 10^6$ rays) using HaloSim \cite{HaloSim} for plate-oriented (gray lines), Parry oriented (blue lines) and horizontal column oriented hexagonal crystals with the aspect ratio set to 2 (a,c) and 0.3 (b,d), as sketched by the insets. The left column is filtered for rays entering and exiting side faces only. The dark grey line assumes tilt angles of $0.1^{\circ}$ and parallel light while the light grey lines assume $1^{\circ}$ and a light source diameter of $0.5^{\circ}$. For high solar elevations the full simulation deviates from the simplified model considerably, see right column. The deviation is smaller for thicker plates.\label{FigHaloSim}}
\end{figure}

%\begin{figure}[tbh]
%\begin{center}\includegraphics [width=1.0\columnwidth]{Fig_PHCircle.pdf} %figure1.png
%  \end{center}\caption{Schematic of the ray paths contributing at various angles to the intensity of the parhelic circle (PHC). The concentric colormaps represent simulated angular RGB spectra at different exposure values, each consecutive being quadrupled in its exposure value. The outmost ring and the inner polar graph is for a for two opposing free faces only. The polar plot has a logarithmic scale spanning about 6 decades. \label{FigLQ}}
%\end{figure}
To compare the angular (azimuthal) positions of the appearing features, such as the limiting PH angles, the effect of a non-zero elevation $e$ of the light source (or sun) relative to the basal prism faces must be taken into account. For non-zero elevations, the deviation can be calculated with the Bravais index of refraction for inclined rays \cite{Bravais1847},
\begin{equation}
n\rightarrow \sqrt{n^2-\sin\left(e\right)^2}/\cos\left(e\right).\label{eq:neff}
\end{equation}
For about $\sim 40^{\circ}$ solar elevation the effective index of refraction of ice reaches the value for a glass prisms at zero light-source elevation.  
A comparison of the results using this simplified approach to a full Monte Carlo ray tracing simulation using HaloSim is shown in Fig.\ \ref{FigHaloSim} for near-zero and $e=20^{\circ}$ elevation. Expectedly, for increasing elevations and thinner plate crystals the simple GOA model starts to deviate significantly from the results obtained. These deviations are due to ray paths entering through the top face $\mathfrak{1}$ (mainly $\mathfrak{1}\textnormal{-}\mathfrak{3}\textnormal{-}\mathfrak{2}$ and $\mathfrak{1}\textnormal{-}\mathfrak{3}\textnormal{-}\mathfrak{8}\textnormal{-}\mathfrak{2}$) which then start to become important. Accordingly, as we restrict ourselves to paths which enter side-faces, many features will be prominent for thick plate oriented crystal displays at very low solar elevations only.

Nonetheless, several features that we discuss are solely due to in-plane ray paths or ones which differ from those only by additional internal reflection from the top and bottom basal faces, leaving the projected ray-path invariant to the in-plane paths we describe \cite{Bravais1847}. This applies to the PH, the Liljequist PH, the $90^{\circ}$-PH and the later discussed round trips. Further, some ray-paths have counterparts on the subparhelic circle \cite{TapeBook}. Again, these differ from the corresponding ray-paths of the PHC only by internal reflections at the top and bottom basal faces. 

%

%Also, some features such as the parhelia and the Liljequist parhelia have their counterpart for randomly oriented columns, producing either the $22^{\circ}$ circular halo or a broad circular backscattering feature around $160^{\circ}$ \cite{Liou1982,Takano1985,Liou1989}.

Finally, we note that imperfections in the crystal shape will not allow certain paths while completely different ones may become possible. Only the external reflection mechanism is mostly insensitive to deviations from perfect hexagonal shapes. Of course, all of the above reservations do not apply to the experiment we describe.

\section{Illuminating a static prism}
\begin{figure}[t]
\centerline{\includegraphics [width=1.0\columnwidth]{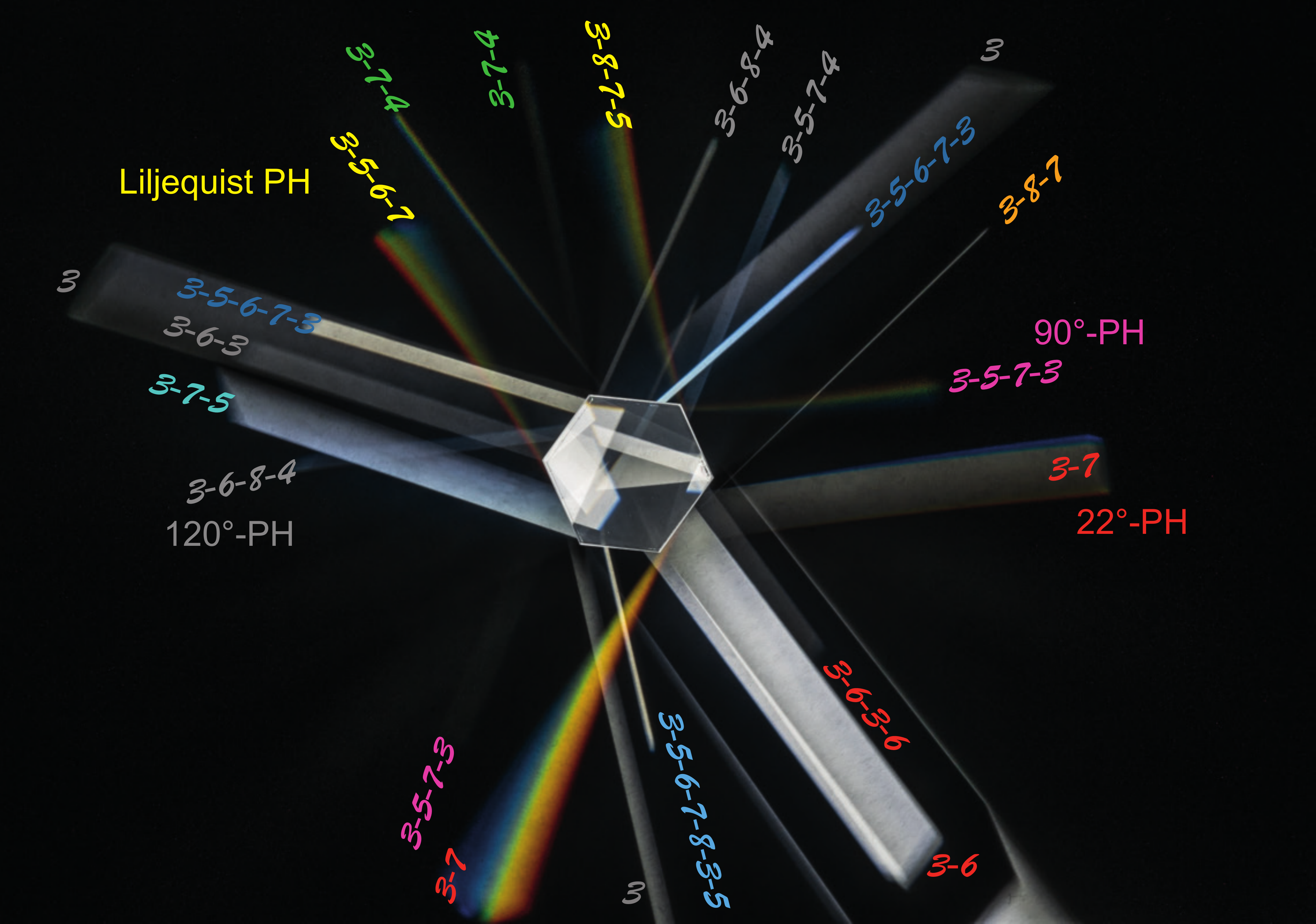}}
\caption{Illuminated BK7 hexagon showing various reflections and refractions as emanating columns. The calligraphic numbers specify the ray-path. The widths indicate the geometrical constraints while the lengths indicate the path complexity. Elevation and orientation were $e\approx 33^{\circ}$ and $\delta \approx 18^{\circ}$, respectively.\label{FigTable}}
\end{figure}

Placing a prism on the table close to a shadow, one may easily observe several emerging beams of light as shown in Fig.\ \ref{FigTable}. Some are the outcome of a paths involving refraction and display a vivid spectrum at certain angles, while others remain white. They correspond to paths with and without net refraction. Still others become blueish at certain angles, signifying earlier onsets of total internal reflections for blue color. 
%The symmetry of the prism causes the pattern to repeat if the hexagon angle exceeds $\delta > 60^{\circ}$. At low sun or light source elevations the lengths (number in parenthesis) of the emerging light columns indicate the corresponding total path lengths through the prism. A shorter column indicates a high number of total internal reflections, cf.\ ray $\mathfrak{3}\textnormal{-}\mathfrak{5}\textnormal{-}\mathfrak{6}\textnormal{-}\mathfrak{7}\textnormal{-}\mathfrak{3}$ and $\mathfrak{3}\textnormal{-}\mathfrak{7}$. 
%
%\begin{figure}[tb]
%\centerline{\includegraphics [width=1.0\columnwidth]{Fig_PrismTables2.pdf}}
%\caption{Sunlit ($e\approx 30^{\circ}$) BK7 hexagon showing various reflections and refractions as emanating columns. The calligraphic numbers specify the ray-path. The widths indicate the geometrical constraints while the lengths (brackets) indicate the path lengths.\label{FigTable}}
%\end{figure}
%

%

\section{Illuminating a rotating prism}
\begin{figure}[b]
\centerline{\includegraphics [width=1.0\columnwidth]{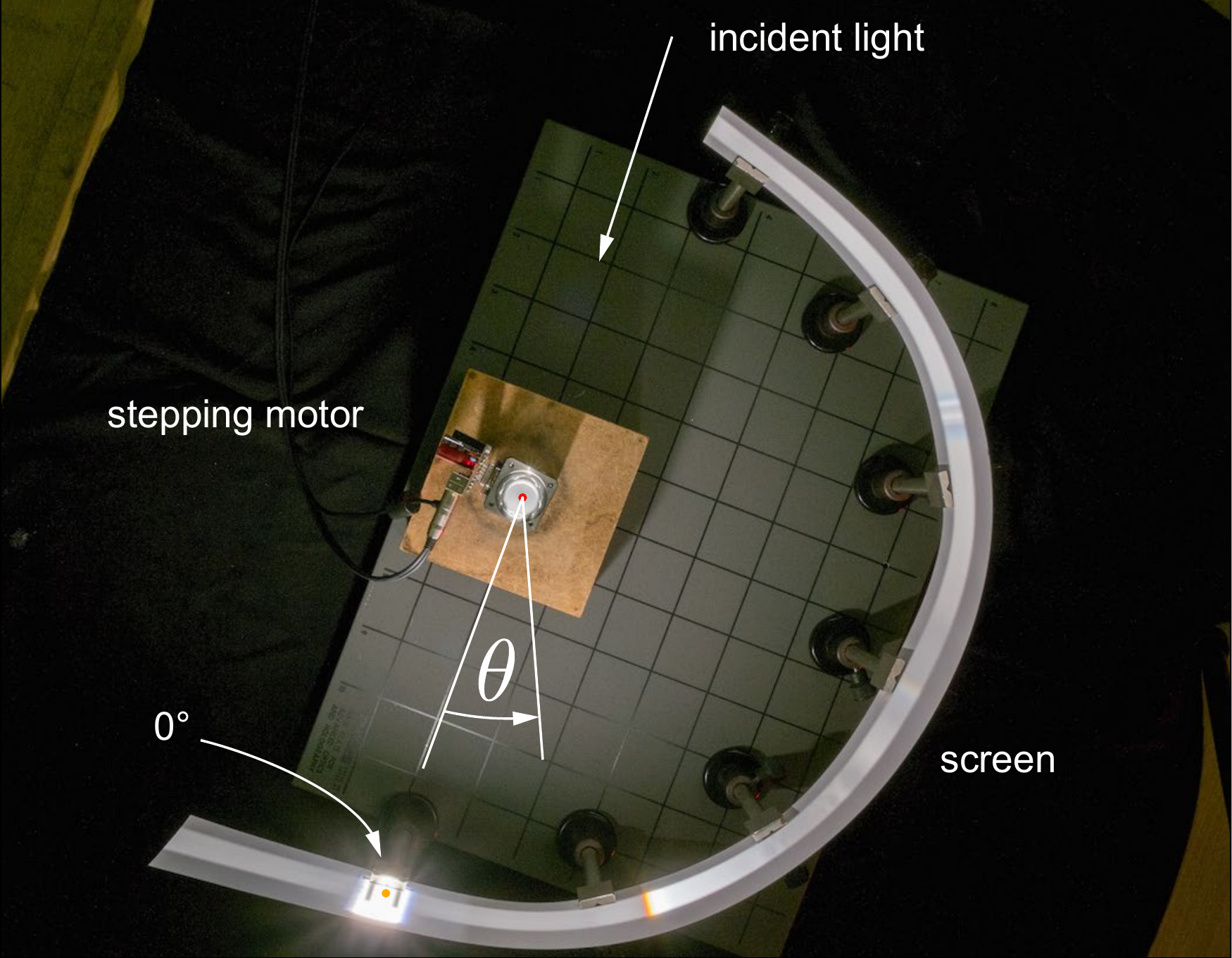}}
\caption{Photograph from above of the illuminated screen for a rotating prism illuminated at $e=0^{\circ}$. A tripod-mounted camera was used to take Raw images in this configuration. Fig.\ \ref{FigIntensity}(b) and Fig.\ \ref{FigIntensityObstr}(b) show the spectra along the screen converted to an angular scale. \label{FigExp}}
\end{figure}

For more quantitative investigations and comparisons with the theory, one may use any camera capable of taking RAW images and long exposure times. To reduce noise, the ISO setting should be set to its minimum value. In our experiment we have used a Fujifilm X-E1 camera with the XF 18-55mm kit zoom-lens, using a maximum exposure time of 30 seconds in the T mode. The RAW files were then converted to linear tiff images using the free software MakeTiff \cite{MakeTiff}. This allows a quantitative assessment of the true intensity recorded by the camera. JPG files already involve a nonlinear processing done by the camera software and are thus only suitable for color images and the analysis of the perception of halo colorings. Using a tripod and rotating the prism using a stepper-motor, very accurate average intensity patterns can be projected on a screen surrounding the prism as shown in Fig.\ \ref{FigExp}. The screen need not be perfectly hemispherical, since a polygonal path may later be used to extract the intensities along with the angular coordinates relative to the forward direction. To compare the recorded images with theory, we first investigated the pattern using a monochromatic blue laser diode operating at wavelength of $\lambda=405\,\rm nm$. Color images using a projector-lamp were recorded next. We also translated wavelength-dependent theoretical intensity distributions into the color images using the spectrum of the light-source, the CIE standard calorimetric observer response function and a typical RGB conversion matrix of the tristimulus values \cite{Kubesh1992}. For more realistic images, the non-linear sRGB values were computed and are shown for comparison along with the experimental data in Fig.\ \ref{FigIntensity} and Fig.\ \ref{FigIntensityObstr}. A similar procedure has previously been applied to study the visibility of tertiary rainbows \cite{Laven2011}.

\section{The individual constituents of the PHC}
In the following sections we discuss the most prominent features in the intensity distribution one by one based on the theoretical ray-paths. The characteristic azimuthal angles will be given for BK7 glass and blue light ($n=1.53$) and for ice ($n=1.31$) to allow both a comparison to our experiment and to features on the natural PHC with the restrictions mentioned before.

\subsection{The $22^{\circ}$ parhelia}
\begin{figure*}[tbh]
\centerline{\includegraphics [width=1.0\textwidth]{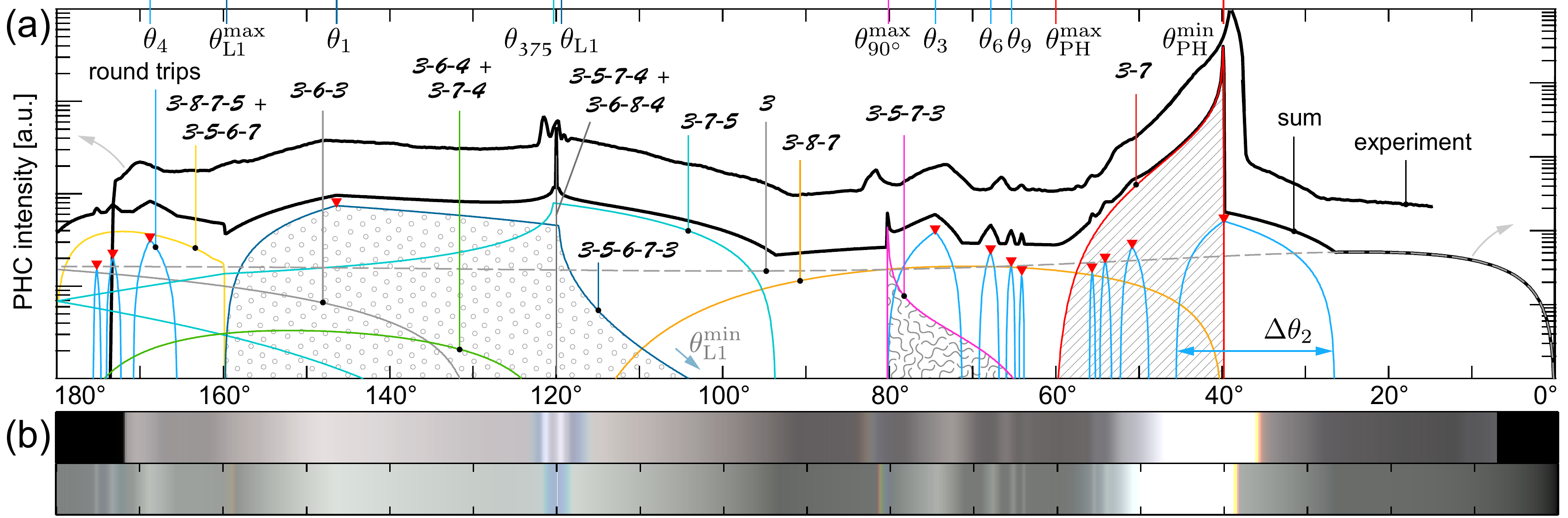}}
\caption{\textbf{(a)} Azimuthal intensity distribution for parallel light incident on the glass hexagon (artificial PHC). Only the paths indicated were considered in the geometrical optics calculation, Eq.\ \eqref{eq:IntGOA}. Individual ray path contributions (coloured thin lines) and their total sum (lower thick black line) are shown, along with the experimental data (upper thick black line). The triangles mark the round-trip ray path azimuths and intensities as computed with Eq.\ \eqref{eq:RT}, section \ref{RT}. Some contributions have been textured to allow easier identification in Figs.\ \ref{FigIntIce} \& \ref{FigIntensityObstr}. The data intensity scale (left) spans one decade less than the theoretical scale (right). \textbf{(b)} Experimental image (top) along the screen (see Fig.\ \ref{FigExp}) and computed image (bottom). White areas correspond to overexposure / saturation.\label{FigIntensity}}
\end{figure*}

According to the ray classification, the ray paths responsible for the parhelia are named $\mathfrak{3}\textnormal{-}\mathfrak{7}$, see Fig.\ \ref{FigPaths}(b). It is readily seen that the halo angle corresponds to the angle of minimum deviation through a $\gamma=60^{\circ}$ prism \cite{Bravais1847,Koennen1983},
\begin{equation}
\sin\left(\frac{\theta_{\rm PH}^{\rm min}+\gamma}{2}\right)=n \sin\left(\frac{\gamma}{2}\right).
\end{equation}
While for the ice halo the result is $\theta_{\rm PH}^{\rm min}\approx 22^{\circ}$, for the BK7 glass prism the angle becomes $\theta_{\rm PH}^{\rm min}\approx 38^{\circ}$, see Fig.\ \ref{FigIntensity}(a). Red colors are refracted more, and hence the PH appear reddish towards the light source, see Fig.\ \ref{FigIntensity}(b). The width $\Delta \theta_{\rm PH} \approx 21^{\circ}$ of the PH can be found width the help of the angle of maximum deviation for grazing incidence on a prism \cite{Bravais1847}, $\theta_{\rm PH}^{\rm max}=\pi/2 - \gamma + \arcsin\left(n\sin\left(\gamma - \arcsin\left(n^{-1}\right)\right)\right)$. In nature, PH appear more narrow since their intensity decays steeply as $1/\sqrt{\theta-\theta_{\rm PH}^{\rm min}}$. The finite angular width of the sun of $0.5^{\circ}$ broadens this divergence to a finite peak of about the same width \cite{Koennen1983}.  This effect is also responsible for the characteristic concave-convex shape in the intensity distribution \cite{White1976,Koennen1983}. The perceived angular extent of the PH is ultimately also determined by the chromatic dispersion of about $1.1^{\circ}$ for ice and $1.6^{\circ}$ for BK7 glass, respectively.
%
%\begin{equation}
%I\left(\theta\right)=\frac{1}{s}\left\{
%\begin{array}{rr}
%\sqrt{\theta-\theta_{\rm PH}+s} & -s<\theta-\theta_{\rm PH}<s\\
%\sqrt{\theta-\theta_{\rm PH}+s}-\sqrt{\theta-\theta_{\rm PH}-s} & \theta-\theta_{\rm PH}>s
%\end{array}\right.
%\end{equation}

\subsection{External and internal reflections}
The simplest explanation of the parhelic circle is that it is caused by external reflections from the side faces of the hexagonal prisms \cite{Bravais1847}, i.e.\ the most simple rays categorized as $\mathfrak{3}$, according to their only face they encounter. An analytical expression for the PHC intensity distribution based from this mechanism can be found in Ref.\ \cite{Lynch1978}. For natural ice halos earlier studies already showed that external reflections contribute negligibly to the PHC \cite{TapeBook}. Instead, the path $\mathfrak{1}\textnormal{-}\mathfrak{3}\textnormal{-}\mathfrak{2}$ is by far the most common contributor at high solar elevations. It consists of a single internal reflection and two compensating refractions. However, for low solar elevations its contribution weakens and different paths such as $\mathfrak{3}\textnormal{-}\mathfrak{8}\textnormal{-}\mathfrak{7}$ become the most prominent contributors \cite{TapeBook}, see also Fig.\ \ref{FigHaloSim}. Only the latter ones are within the scope of the simple model discussed here. A brief examination of the experimental as well as calculated data leads to the conclusion, that only within the near-forward direction between the light source and the PH external reflections contribute appreciably.

\subsection{The $90^{\circ}$ parhelia}
In laboratory experiments the so-called $90^{\circ}$-parhelia were observed and explained \cite{Bravais1847,Tammer1998} by the ray-path $\mathfrak{3}\textnormal{-}\mathfrak{5}\textnormal{-}\mathfrak{7}\textnormal{-}\mathfrak{3}$, see Fig.\ \ref{FigPaths}(b). Already the ray-path suggests its close similarity to the $22^{\circ}$ parhelia. The consecutive internal reflections by two faces making an angle $\gamma=60^{\circ}$ results in an additional deflection by $2\left(\pi-\gamma\right)$. Since the total deviation $\psi$ exceeds $180^{\circ}$, the actual observed angle towards the light source will be $2\pi-\psi$. Accordingly, the analysis of the angle of deflection yields an angle of minimum deviation (but maximum angular distance from the light source) at $\theta_{90^{\circ}}^{\rm max}=2\pi/3-\theta_{\rm PH}^{\rm min}=2\arccos\left(n/2\right)$, with an inverted color sequence for this halo phenomenon as compared to the common PH, see Fig.\ \ref{FigIntensity}(b). Notably, the ray path is similar to the secondary rainbow, including the colour sequence \cite{Walker1975}. For ice the halo would appear at about $\theta_{90^{\circ}}^{\rm max}\approx 98^{\circ}$, while for BK7 the angle is $\theta_{90^{\circ}}^{\rm max}\approx 80^{\circ}$. However, the fact that the intensity of the $90^{\circ}$-PH is approximately 3 orders of magnitude fainter than the $22^{\circ}$ PH explains why it has not yet been observed in nature and makes future observation difficult at best. In the glass prism experiment, where the otherwise dominant path $\mathfrak{1}\textnormal{-}\mathfrak{3}\textnormal{-}\mathfrak{2}$ is absent, the halo can be observed clearly. Following an approach originally due to A.\ Bravais \cite{Bravais1847}, obstructing all but two opposing faces blocks out the intense parhelia. Then, the $90^{\circ}$-PH will show up as the brightest full-spectrum halo. It has been speculated before \cite{Bravais1847} that Hevel's curious circular halo from 1661 could be explained by this ray path, although full 3D ray-tracing simulations by W.\ Tape could not support this conjecture \cite[Display 11-1]{TapeBook}. Lastly, an even weaker second order $90^{\circ}$-PH has been observed by Bravais using a triangular prism \cite{Bravais1847}. Its ray path corresponds to $\mathfrak{3}\textnormal{-}\mathfrak{7}\textnormal{-}\mathfrak{5}\textnormal{-}\mathfrak{3}\textnormal{-}\mathfrak{7}\textnormal{-}\mathfrak{5}$ with a minimum deflection angle of $\theta_{90^{\circ},2}^{\rm min}=2\pi/3+\theta_{\rm PH}^{\rm min}$ with the same color sequence as the $22^{\circ}$-PH.

\begin{figure*}[tbh]
\centerline{\includegraphics [width=1.0\textwidth]{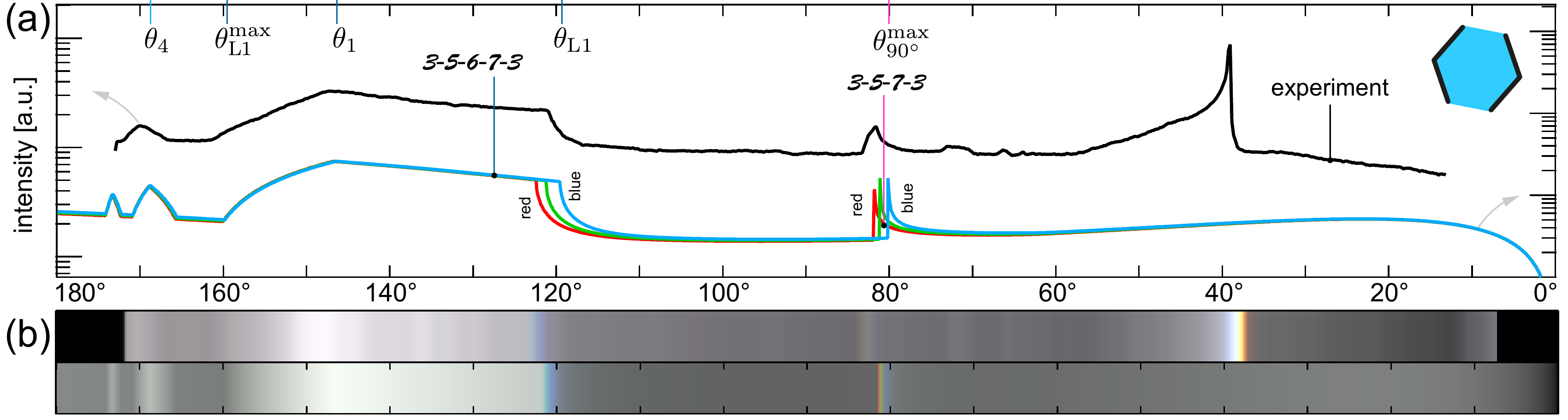}}
\caption{\textbf{(a)} Azimuthal intensity distribution when all but two opposing faces of the prism are masked (see inset sketch). The experimental data (black) as well as the geometrical optics calculation for $n=\left\{1.53,1.52,1.51\right\}$ (blue, green, red). The residual PH around $39^{\circ}$ is due to imperfect masking. \textbf{(b)} Experimental image (top) along the screen (see Fig.\ \ref{FigExp}) and computed image (bottom) for the cardboard-covered prism.\label{FigIntensityObstr}}
\end{figure*}

\subsection{The $120^{\circ}$ parhelia}
The natural white $120^{\circ}$ parhelion at large solar elevations is predominantly caused by $\mathfrak{1}\textnormal{-}\mathfrak{3}\textnormal{-}\mathfrak{8}\textnormal{-}\mathfrak{2}$ rays \cite{TapeBook}, cf.\ Fig.\ \ref{FigHaloSim}(d). This path involves two internal reflections by the adjacent faces $\mathfrak{3}$ and $\mathfrak{8}$, and thus cause a constant deflection by $120^{\circ}$. The ray enters and leaves through the top and bottom basal faces, respectively. At low solar elevations, the former path looses intensity and a very different path, namely the ray $\mathfrak{3}\textnormal{-}\mathfrak{5}\textnormal{-}\mathfrak{7}\textnormal{-}\mathfrak{4}$, becomes important \cite{Bravais1847,TapeBook}. This path resembles the one responsible for the $90^{\circ}$-PH shown in Fig.\ \ref{FigPaths}(b), but instead of exiting through the entry face $\mathfrak{3}$ it exits through the adjacent one $\mathfrak{4}$. The internal angles for the entry and exit are identical such that no net-refraction occurs. Accordingly, one finds that such a ray experiences an orientation- and color-independent constant total deflection by $\psi=4\pi/3$, corresponding to an azimuth of $\theta_{120^{\circ}}=2\pi-\psi\equiv 120^{\circ}$. The reverse path $\mathfrak{3}\textnormal{-}\mathfrak{6}\textnormal{-}\mathfrak{8}\textnormal{-}\mathfrak{4}$ is also possible. In the experiment, upon rotation of the prism, one observes a non-moving bright spot. Although the deflection remains constant, a homogeneously illuminated prism will redirect a pencil of rays of different width depending on the orientation of the prism. If a screen is located far enough from the prism, this halo shrinks to the angular extent or divergence of the light-source. This is similar to the natural counterpart which should have an extent determined by the angular width of the sun and the orientational ordering of the plate crystals. In the experiment, at closer distances the angular extent of this feature becomes limited by the size of the prism. Accordingly, the two bright stripes in the experimentally recorded color spectra and the two peaks in the monochromatic intensity distribution directly indicate the average of the effective areas of the incidence beam that are intercepted and redirected into this direction, see Fig.\ \ref{FigIntensity}(b).

\subsection{The Liljequist parhelion} %\cite{LiljequistReport,ursa495}
\begin{figure}[t!!]
\centerline{\includegraphics [width=1.0\columnwidth]{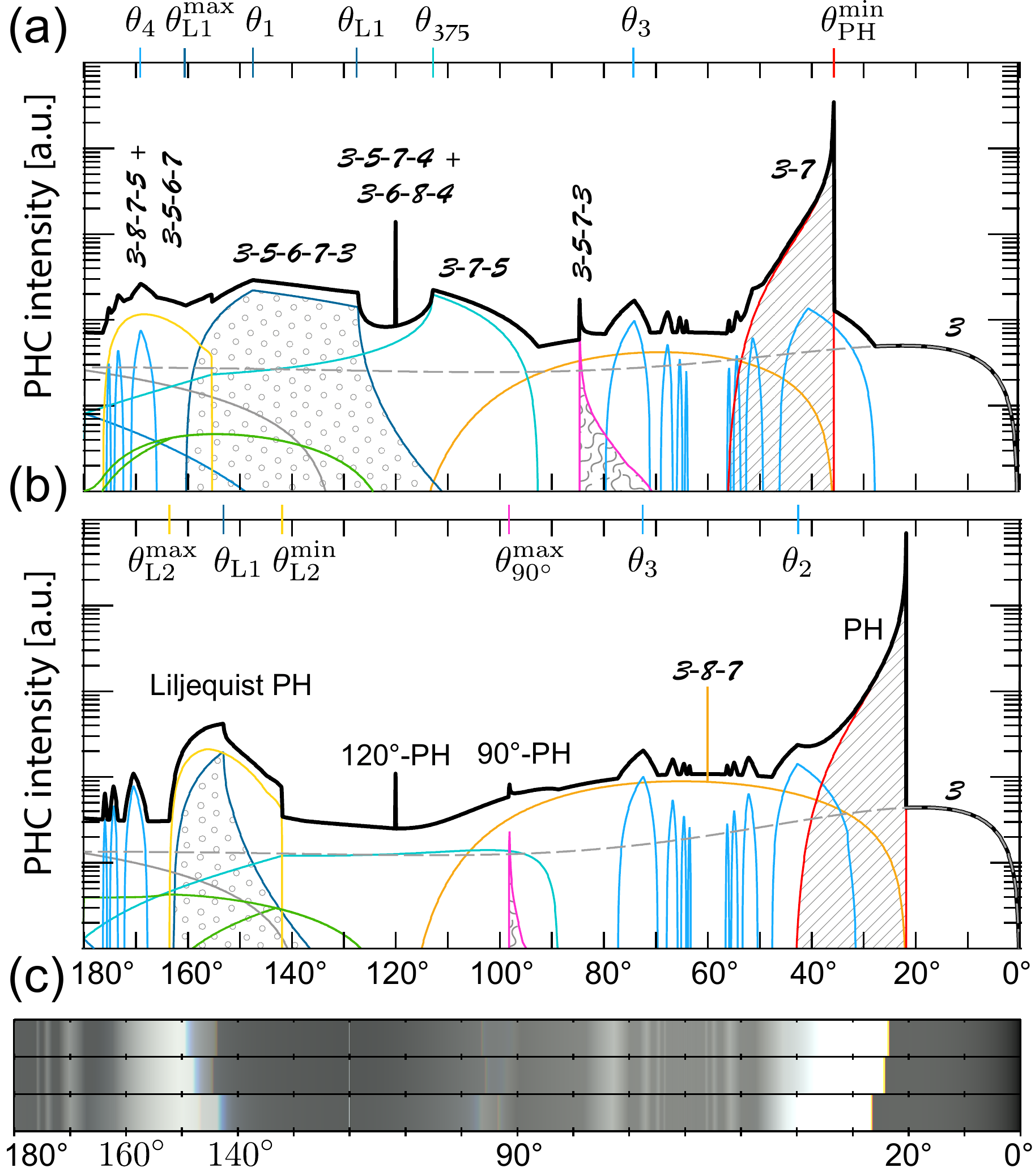}}
\caption{Intensity plot as in Fig.\ \ref{FigIntensity}(a), but for \textbf{(a)} acrylic glass, $n=1.48$ as in Fig.\ 14 of Ref.\ \cite{Tammer1998} and \textbf{(b)} ice, $n=1.31$ (cf.\ Refs.\ \cite{Liou1989,Macke1996,Borovoi2007,Flatau2014}). \textbf{(c)} Coloring for ice (PH are overexposed) at different solar elevations $e=\left\{0^{\circ},10^{\circ},20^{\circ}\right\}$ (top to bottom). Only in-plane paths were considered, i.e.\ not the blue spot phenomenon.\label{FigIntIce}}
\end{figure}

First observed by G\"osta Hjalmar Liljequist in 1951 in Antarctica \cite{LiljequistReport}, the Liljequist parhelion denotes the intensity feature associated with rays that are totally internally reflected twice. It was first simulated by E.\ Tr\"ankle and R.\ G.\ Greenler in 1986 \cite{Greenler1986} and then explained by W.\ Tape in 1994 \cite{TapeBook}. It was subsequently investigated regarding its coloring with Monte-Carlo methods by J.\ Moilanen in 1996 \cite{ursa596}. Unambiguous photographs of this rare halo can be found for O.\ SŠlevŠ's Rovaniemi display (28 Oct.\ 2012) and for the spotlight experiments done by M.\ Riikonen (Rovaniemi, 7/8th Dec.\ 2008). Also in the book of W.\ Tape the subparhelic counterpart is shown \cite[Display 7-1]{TapeBook}. For ice, this feature appears at about $27^{\circ}$ to both sides of the anthelic point, cf.\ Fig.\ \ref{FigIntIce}(b). Primarily two ray paths contribute to this halo, namely the $\mathfrak{3}\textnormal{-}\mathfrak{5}\textnormal{-}\mathfrak{6}\textnormal{-}\mathfrak{7}\textnormal{-}\mathfrak{3}$ ray and the $\mathfrak{3}\textnormal{-}\mathfrak{8}\textnormal{-}\mathfrak{7}\textnormal{-}\mathfrak{5}$ (and its reverse) ray. Figure \ref{FigPaths} shows photographs of both paths in (e) and (c), respectively. For BK7 glass both paths cause PHC intensities at clearly separate azimuths. Considering the former ray path first, the angular extent and position is determined by two effects. A steep decrease in intensity occurs at an angle corresponding to the deflection angle of this ray when the identical internal reflections at interfaces $\mathfrak{5}$ and $\mathfrak{7}$ become total internal reflections, cf.\ the schematic in Fig.\ \ref{FigArea}. The corresponding azimuth when this happens is given by
\begin{equation}
\theta_{{\rm L}1}=2\arccos\left(n\cos\left(\frac{\pi}{6} + \alpha_{\rm TIR}\right)\right). %\textnormal{arcsin}\left(\frac{1}{n}\right)
\end{equation}
For BK7 glass the angle is $\theta_{{\rm L}1}\approx 120^{\circ}$. For ice the azimuth is $\theta_{{\rm L}1}\approx 153^{\circ}$ and marks the coordinate of highest intensity of the Liljequist PH. For blue light the angle is smaller by about $2^{\circ}$ for both ice and glass, such that the halo begins with a transition from blue towards the sun over cyan to white and Fig.\ \ref{FigIntIce}(c). The coloring is therefore not mediated by a net refraction of the rays which emerge from the prism, in contrast to the $22^{\circ}$- and $90^{\circ}$-PH. Experimentally, the effect is best seen when separated from other contributors in this "busy azimuthal domain", see Fig.\ \ref{FigIntensity}(a),(b) around $120^{\circ}$. The obstruction of all but two opposing crystal faces allows an individual investigation of this peculiar halo contribution, see Fig.\ \ref{FigIntensityObstr}(a),(b). In particular, this restriction of partaking prism faces also eliminates the $120^{\circ}$-PH and the overlapping $\mathfrak{3}\textnormal{-}\mathfrak{7}\textnormal{-}\mathfrak{5}$ contribution.
The maximum angle of this halo is determined by the geometrical constraint that the exit ray must still hit the entrance face. Successive application of the law of sines may be used to find the corresponding entrance ray positions admissable, whereby the azimuthal limits are found as
\begin{equation}
\left[\theta_{{\rm L}1}^{\rm min},\theta_{{\rm L}1}^{\rm max}\right]=\left[2\arccos\left(\frac{n}{2}\right),2\arccos\left(\frac{n}{2\sqrt{19}}\right)\right],
\end{equation}
such that the azimuthal beginning of this feature coincides with the boundary of the $90^{\circ}$-PH, i.e.\ $\theta_{{\rm L}1}^{\rm min}=\theta_{90^{\circ}}^{\rm max}$. For ice the azimuthal range is $\left[98^{\circ},163^{\circ}\right]$, whereas for BK7 glass this amounts to a range of $\left[80^{\circ},160^{\circ}\right]$, see Fig.\ \ref{FigIntensity}(a). The decay towards the low angle side at $\theta_{{\rm L}1}^{\rm min}$ occurs about three decades below the plot-range and is thus not seen in the figure.

The second contributor is the $\mathfrak{3}\textnormal{-}\mathfrak{8}\textnormal{-}\mathfrak{7}\textnormal{-}\mathfrak{5}$ ray path.
Its azimuthal limits are found to be
\begin{eqnarray}
\theta_{{\rm L}2}^{\rm max}&=&\frac{5\pi}{6} + \arcsin\left(n\cos\left(\frac{\pi}{6}+\alpha_{\rm TIR}\right)\right),\nonumber\\
\theta_{{\rm L}2}^{\rm min}&=&\frac{\pi}{3} + 2\arcsin\left(\frac{n}{2}\right).
\end{eqnarray}
For BK7 glass this corresponds to an azimuthal range of $\left[160^{\circ},180^{\circ}\right]$, whereas for ice this corresponds to a range of $\left[142^{\circ},163^{\circ}\right]$, see Fig.\ \ref{FigIntIce}(a). For ice crystals, both ray paths contribute almost equally to the Liljequist PH, while its visible angular width should be determined by the latter contribution, i.e.\ $\sim 21^{\circ}$ at almost zero solar elevation. Figure \ref{FigIntIce}(c) shows simulations of the coloring of the Liljequist parhelia expected for low solar elevations and match the photographs from the spotlight experiments by M.\ Riikonen.

\subsection{The blue spot}
Recently, the PHC blue spot phenomenon was explained \cite{BlueSplot2001} by the properties of the ray path $\mathfrak{1}\textnormal{-}\mathfrak{3}\textnormal{-}\mathfrak{2}$. A wavelength-dependent total internal reflection at interface $\mathfrak{3}$ is responsible for the blue spot occurring at an azimuth $\theta_{\mathfrak{1}\mathfrak{3}\mathfrak{2}}=2\arcsin\left(n\cos\left(\arcsin\left(1/n\right)\right)/\cos\left(e\right)\right)$. In this expression, the index of refraction rather than the Bravais index of refraction is to be used. In Fig.\ \ref{FigHaloSim}(d) this angle corresponds to the drop-off of the intensity near $128^{\circ}$, cf.\ Fig.\ 2 and 3 in Ref.\ \cite{BlueSplot2001}. Possibly, the unaccounted excess azimuthal width of the blue spot can be attributed to the ray path $\mathfrak{3}\textnormal{-}\mathfrak{5}\textnormal{-}\mathfrak{6}\textnormal{-}\mathfrak{7}\textnormal{-}\mathfrak{3}$, encountered in the previous study of the Liljequist PH. Its blueish edge color at $\theta_{{\rm L}1}$ overlaps with the blue spot for solar elevations $e\approx 25^{\circ}$. This coincides with the conditions for which the majority of the photographic evidence exists \cite{BlueSplot2001}. Whether a contributor or not, the same mechanism of wavelength-dependent total internal reflection causes the blue spot observed in the experiment. It is therefore an illustrative explanation for the blue spot ice halo phenomenon on the PHC.

In the table top-experiment, the ray path $\mathfrak{3}\textnormal{-}\mathfrak{7}\textnormal{-}\mathfrak{5}$ also shows a blueish edge. This is due to the occurrence of total internal reflection at the second interface $\mathfrak{7}$, leading to a steep drop-off of the intensity at a characteristic azimuthal angle $\theta_{\mathfrak{3}\mathfrak{7}\mathfrak{5}}$:
\begin{equation}
\theta_{\mathfrak{3}\mathfrak{7}\mathfrak{5}} = \frac{\pi}{3} + 2\arcsin\left(n \cos\left(\frac{\pi}{6} + \alpha_{\rm TIR}\right)\right).
\end{equation}
While this path only contributes little to the natural ice halo, it shows up clearly for glass, cf.\ Fig.\ \ref{FigIntensity}. For BK7 glass it appears at $120^{\circ}$, shifting towards smaller angles for smaller refraction indices, cf.\ Fig.\ \ref{FigIntIce}(a) for acrylic glass with $n=1.46$. Since $\theta_{{\rm L}1}$ and $\theta_{\mathfrak{3}\mathfrak{7}\mathfrak{5}}$ coincide for BK7 glass and occur at the position of the $120^{\circ}$-PH, a complex blue/white feature is seen in our experiment with $e=0^{\circ}$, see Fig.\ \ref{FigIntensity}(b).

\subsection{Round trips\label{RT}}
If the hexagon is oriented such that a face is almost normal to the incidence beam ($\delta$ small, cf.\ Fig.\ \ref{FigArea}), round trips can be observed for rays impinging on the face close to its end. The Liljequist PH ray path $\mathfrak{3}\textnormal{-}\mathfrak{5}\textnormal{-}\mathfrak{6}\textnormal{-}\mathfrak{7}\textnormal{-}\mathfrak{3}$ constitutes the round trip of order $i=1$, and is shown in Fig.\ \ref{FigPaths}(e). The construction of the $i$-th order round trip is now recursively defined as follows: The exiting face of the preceding round trip is replaced by two total internal reflections at the very same face and its predecessor, followed by exiting through the face opposite to the predecessor. Accordingly, the second order $i=2$ round trip ray path is $\mathfrak{3}\textnormal{-}\mathfrak{5}\textnormal{-}\mathfrak{6}\textnormal{-}\mathfrak{7}\textnormal{-}\mathfrak{8}\textnormal{-}\mathfrak{3}\textnormal{-}\mathfrak{5}$ as shown in Fig.\ \ref{FigPaths}(f). Photographs of the first seven round trips can be found in the supplementary information. Each increase in the round trip order increases the total deflection angle by $120^{\circ}$. Staring with the first order round trip, the maxima occur for orientations around $\delta_i = \arcsin(n/\sqrt{1+3x^2})$ and cause corresponding deviations of
\begin{equation}
\theta_{i}=\frac{\pi}{3}x+2\arcsin\left(\frac{n}{\sqrt{1+3x^2}}\right) \,{\rm mod}\, 2\pi,\label{eq:RT}
\end{equation}
with $x=2i+1$ and $i=1,2\dots$, such that they occur at azimuths of $\theta_{i}$ or $2\pi-\theta_{i}$ when $\theta_{i}>\pi$. For zero elevation, BK7 glass and blue light ($n=1.53$) these maxima occur at $\theta_{1,2,\dots}=146^{\circ}$, $40^{\circ}$, $74^{\circ}$, $169^{\circ}$, $51^{\circ}$, $68^{\circ}$, $173^{\circ}$, $54^{\circ}$, $65^{\circ}$, $175^{\circ}$. The sensitivity of the deviations in Eq.\ \eqref{eq:RT} on the refractive index (and thus also on elevation) is weak, such that these features appear colorless. Their intensity drops as $1/x$ such that most of them could be observed in the experiment, see Fig.\ \ref{FigIntensity}(a),(b). The reason for weak decay are the total internal reflections occurring at all but the entry and exit faces. The drop almost exclusively stems from the geometrical constraint, similar to Fig.\ \ref{FigArea}. These constraints were also used to arrive at Eq.\ \eqref{eq:RT}. The triangular solid markers in Fig.\ \ref{FigIntensity}(a) show the angles computed with Eq.\ \eqref{eq:RT} and the scaling $\propto 1/x$. Starting with $i=2$, the widths of these peaks narrow down as
\begin{equation}
\sin\left(\Delta \theta_{i}\right) \approx \frac{8 n}{\sqrt{3}}x^{-2} + \frac{4n}{\sqrt{3}}\left[7+\frac{n^2}{3}\right] x^{-4}\label{eq:RTwidths},
\end{equation}
which is accurate to within less than 1\%. A previous experiment also showed the first three round-trips \cite{Tammer1998}, although they are not discussed and higher orders are absent, presumably due to the use of a low-quality hexagonal prism. As this phenomenon requires near-perfect hexagons, it is doubtful whether any indication of it should be expected in nature. If such ice halo should be observable, the features near $\theta_2=43^{\circ}$, $\theta_3=72^{\circ}$ and $\theta_4=170^{\circ}$ for very low solar elevations are the best candidates, see Fig.\ \ref{FigIntIce}(a). The appearance would indicate almost perfect hexagonal ice crystals. A similar indicative ability has been identified by M.\ Riikonen in a blog article \cite{MarkoRiikonenBlog}. He found that imperfect hexagons would result in the disappearance of the Liljequist PH.
Finally, the partial retroreflection at the exit-face of these rays (cf. Fig.\ \ref{FigPaths}(e,f)) cause further faint features at constant azimuths of $120^{\circ}$ and $0^{\circ}$.

\section{Outlook and Summary}
\begin{figure}[bt]
\centerline{\includegraphics [width=1.0\columnwidth]{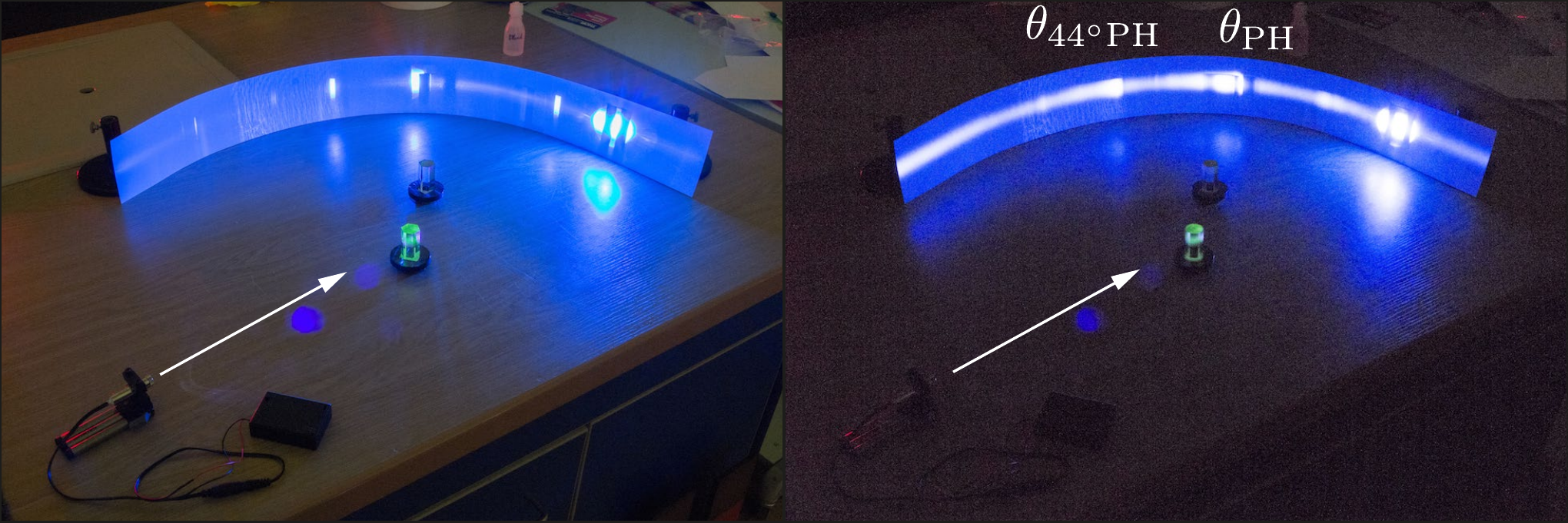}}
\caption{Experimental setup for the demonstration of the mechanism causing the $44^{\circ}$ parhelia. The left and right image shows the situation without and with rotation, respectively. The prisms have been mounted on axial ball bearings which sustained rotation for a few seconds. \label{Fig44}}
\end{figure}

Using two prisms, the mechanism for the $44^{\circ}$ parhelia may be investigated. Placing a second prism close to the first prism at an angle corresponding to the PH angle $\theta_{\rm PH}$, the successive deflection through both prisms leads to a feature at twice the minimum deflection angle $\theta_{44^{\circ}\rm PH}^{\rm min}=2\theta_{\rm PH}^{\rm min}$, i.e.\ the $44^{\circ}$ parhelia. For BK7 glass, this halo appears at $\approx 80^{\circ}$ as clearly visible in the experiment, see Fig.\ \ref{Fig44}. Also Parry arcs may be studied with a single rotation axis \cite{Tammer1998}. Other halo phenomena such as the tangential arcs may be observed with more complex setups (\cite{Tammer1998,Greenler2003}, cf.\ also the work of Michael Gro\ss mann, Arbeitskreis Meteore e.V.). Unfortunately, neither the spectacular circumhorizontal arc nor the circumzenithal arc can be investigated using a glass prism \cite{Tammer1998}. The responsible ray paths, $\mathfrak{1}\textnormal{-}\mathfrak{3}$ and $\mathfrak{3}\textnormal{-}\mathfrak{2}$, respectively, correspond to refraction through a $\gamma=90^{\circ}$ prism. However, an outgoing ray is prevented by total internal reflection at the second interface for any material with a refractive index $n>\sqrt{2}$, which is the case for all practical prism materials. Making an impression of the glass prism with some moulding material, a hexagonal mould for ice crystals may be produced. The use of pre-boiled water prevents the formation of bubbles during freezing and thereby allows the construction of clear ice crystal models. Alternatively, the use of water-filled prisms has been proposed by A.\ Bravais \cite{Bravais1847}, who used a spinning triangular prism. It should be most interesting to think about further visualisation and demonstration possibilities in this field of atmospheric optics.

In summary, we have presented a theoretical approach to efficiently obtain the intensity distribution of the natural parhelic circle for low solar elevations as well as for artificially generated parhelic circles obtained in rotating prism halo demonstration experiments. Based on a ray-by-ray analysis in geometrical optics, we gave explicit expressions for several characteristic angles on the PHC for plate-oriented crystals. Particularly, the two superposing constituents of the Liljequist parhelia were shown to be connected to two corresponding but separable features in the glass-prism experiment.

\acknowledgements
We acknowledge financial support by Prof.\ Frank Cichos and thank Axel M\"arcker for help with experimental equipment from the lecture hall collection. We also thank two anonymous referees who improved this manuscript with their comments.

%\begin{thebibliography}{99}
%% Do not include separate BibTeX files; if BibTeX is used,
%% paste the output (contents of .bbl file) here.
%\bibitem{revtex-au} \url{https://authors.aps.org/revtex4/}.
%\bibitem{osastyle} \url{http://www.opticsinfobase.org/submit/style/jrnls_style.cfm}.

\bibliographystyle{osajnl}
\bibliography{PhysTeacher}

\end{document}